\documentstyle{article}
\begin{document}
\title{A~Large~Solid~Angle~Study~of~Pion Absorption~on~$^3$He}
\author{
T. Alteholz,$^{\it c}$
D. Androi\'c,$^{\it l}$
G. Backenstoss,$^{\it a}$
D. Bosnar,$^{\it l}$ \\
H. Breuer,$^{\it e}$
A. Brkovi\'{c},$^{\it l}$
H. D\"obbeling,$^{\it k}$
T. Dooling,$^{\it j}$ \\
W. Fong,$^{\it f}$
M. Furi\'c,$^{\it l}$
P.A.M. Gram,$^{\it d}$
N.K. Gregory,$^{\it f}$ \\
J.P. Haas,$^{\it h}$
A. Hoffart,$^{\it c}$
C.H.Q. Ingram,$^{\it k}$
A. Klein,$^{\it j}$ \\
K. Koch,$^{\it k}$
J. K\"ohler,$^{\it a}$
B. Kotl\'{\i}nski,$^{\it k}$
M. Kroedel,$^{\it a}$ \\
G. Kyle,$^{\it h}$
A. Lehmann,$^{\it a}$
Z.N. Lin,$^{\it h}$
G. Mahl,$^{\it k}$ \\
A.O. Mateos,$^{\it f}$
K. Michaelian,$^{\it k}$
S. Mukhopadhyay,$^{\it h}$ \\
T. Petkovi\'{c},$^{\it l}$
R.P. Redwine,$^{\it f}$
D. Rowntree,$^{\it f}$ \\
R. Schumacher,$^{\it b}$
U. Sennhauser,$^{\it k}$
N. \v{S}imi\v{c}evi\'{c},$^{\it f}$ \\
F.D. Smit,$^{\it g}$
G. van der Steenhoven,$^{\it i}$
D.R. Tieger,$^{\it f}$ \\
R. Trezeciak,$^{\it c}$
H. Ullrich,$^{\it c}$
M. Wang,$^{\it h}$
M.H. Wang,$^{\it h}$ \\
H.J. Weyer,$^{\it a,k}$
M. Wildi,$^{\it a}$
K.E. Wilson$^{\it f}$\\
\\
\\
(LADS collaboration)\\
\\
\\
$^{\it a}$ University of Basel, CH-4056 Basel, Switzerland\\
$^{\it b}$ Carnegie-Mellon University, Pittsburgh PA 15213, USA\\
$^{\it c}$ University of Karlsruhe, D-7500 Karlsruhe, Germany\\
$^{\it d}$ LAMPF, Los Alamos NM 87545, USA\\
$^{\it e}$ University of Maryland, College Park MD 20742, USA\\
$^{\it f}$ Massachusetts Institute of Technology, Cambridge MA 02139, USA\\
$^{\it g}$ National Accelerator Center, Faure 7131, South Africa\\
$^{\it h}$ New Mexico State University, Las Cruces NM 88003, USA\\
$^{\it i}$ NIKHEF-K, NL-1009 DB Amsterdam, The Netherlands\\
$^{\it j}$ Old Dominion University, Norfolk VA 23529, USA\\
$^{\it k}$ Paul Scherrer Institute, CH-5232 Villigen PSI, Switzerland\\
$^{\it l}$ University of Zagreb, HR-41001 Zagreb, Croatia\\
}

\date{April 6, 1994}
\maketitle
\begin{abstract}
Measurements have been made of $\pi^+$ absorption on $^3$He at
T$_{\pi^+}$ = 118, 162,
and 239~MeV using the Large Acceptance Detector System (LADS).
The nearly 4$\pi$ solid angle coverage of this detector minimizes
uncertainties associated with extrapolations over unmeasured
regions of phase space.
The total absorption cross section
is reported.  In addition, the total cross
section is divided into components in which only
two or all three nucleons play a
significant role in the process.
These are the first direct measurements of the total and three nucleon
absorption cross sections.
\end{abstract}

\medskip
Nuclear pion absorption at energies near that
of the $\Delta$(1232)-resonance has been the
focus of much attention in recent years~\cite{Ashery86,Weyer90,Ingram92}.
The two-nucleon process, primarily
absorption on T~=~0 ``deuteron-like'' pairs,
is experimentally well established and was
initially expected to dominate the absorption process on all
nuclei.  However, more detailed studies
have indicated that multi-nucleon
mechanisms play a significant role.
``Multi-nucleon mechanisms'' refer to absorption processes
that involve more than two nucleons, none of which acts simply as a
spectator.  It is sometimes appropriate to separate further
those events which are characterized by signatures of known
subprocesses, such as initial and final state interactions.

Indications of multi-nucleon absorption modes have been seen in
discrepancies between the measured total absorption cross section
and the measured
two-nucleon absorption cross section on medium-heavy
nuclei \cite{Ashery81,Altman83,Burger86,Schumacher88,Hyman93}.
These modes
are most easily studied in detail in light nuclei,
especially $^3$He and $^4$He, where
the effects of the addition of one or two nucleons to the absorbing system
can be isolated.

Previous absorption experiments on these targets have shown the
existence of multi-nucleon absorption modes
\cite{Backen85,Backen88,Smith89,Stein90,Mukhop91,Weber91A,Weber91C,Adimi92}.
However, they all suffered from some limitations in angular coverage
or resolution,
kinematic definition, or statistical precision for multi-particle final states.
The Large Acceptance Detector System (LADS) was constructed at the Paul
Scherrer Institute (PSI) to examine pion
absorption reactions with an emphasis on multi-nucleon final states.
LADS was designed to cover as much solid angle as possible with good
energy and angular resolution,
with the ability to detect neutrons in the final state, and with
enough granularity to distinguish the outgoing nucleons.
This letter reports the total absorption cross section for $\pi^+$ absorption
on $^3$He at three energies around the $\Delta$-resonance.
The partial cross sections corresponding to the
cases in which only two or all three nucleons play a significant role in
the absorption process are also reported.

The LADS detector (Fig.~1) \cite{Backenstoss91} consists of a
cylinder of 28 dE-E-E plastic
scintillator sectors, each 1.6~m in active length
and viewed by photomultiplier
tubes at each end.  The dEs are 4.5~mm thick.
The E layers can stop 200~MeV normally incident
protons, with an energy resolution of 3\% FWHM.  The ends of the cylinder
are closed by ``endcaps'', each consisting of 14 dE-E plastic scintillator
sectors.
More than 98\% of $4\pi$ solid angle is covered by the scintillator.

For charged particles, trajectory information is provided by two
concentric cylindrical multi-wire proportional chambers (MWPCs).
The inner chamber
has a radius of 6.4~cm and covers the
entire solid angle subtended by the detector;
the outer chamber has a radius of 28~cm and covers the same
region as the cylinder scintillators.  Combined, these give angular resolutions
for charged particles of about 10~mr FWHM.

The target is a high pressure (95~bar) gas cylinder 25.7~cm in length with
a 2~cm radius.  The walls are constructed of 0.5~mm thick carbon
fiber composite with a 30~$\mu$m copper-silver lining.
Energy loss in the walls
of the target and in the MWPCs dominates the threshold of the
detector system, which is below 20~MeV for
protons.

The data were collected in the $\pi$M1 area at PSI.  The $\pi^{+}$ beam was
defined using plastic scintillator detectors that counted individual pions.
The defined beam rate was typically $10^{5}$/s out of a total incident flux
of about $3 \times 10^6$/s.
Events were classified according to the
number of charged and neutral particles detected in LADS in coincidence with a
defined beam pion, and a fraction of each possible combination was
written to tape.
Different final states could be emphasized
by varying this sampling fraction.

There is only one significant final state for the absorption of
a $\pi^+$ by $^3$He: three unbound protons.  Identification of the absorption
reaction depends only on determining that there are no pions in the final
state.
The disappearance of the incident pion was determined by the requirement
that more energy be deposited in the scintillators than the kinetic energy
of the incident pion, assisted by
energy {\it vs.} dE/dx and energy {\it vs.} time-of-flight
particle identification techniques.
The vertex of each event measured by the MWPCs was used to reject
absorption events originating in the target's entrance and exit windows.
Frequencies for the detection of two or of three
protons from an absorption event were then determined.

It was necessary to correct these experimental frequencies
for events lost due to reactions of the protons in the plastic
scintillator (3-18\%);
for wire chamber inefficiency (4-10\%); for background events (4-15\%); and for
events in which scattered pions which reacted in the scintillator
were classified as protons (0.2-3.3\%).
The corrections vary with the incident pion energy over the range indicated
and were all determined from the data.
The measurement of reaction losses is described in the next
paragraph as an example.

To correct the yield when only two protons were detected, data were used
from absorption on deuterium since the energy and angular distributions
are similar to those from the two-nucleon absorption process on $^3$He.
Absorption events were selected from their angular correlations
by the wire chambers, independently of the scintillator energy information.
The fraction of these events lost while passing through the normal analysis
chain determined the correction to the $^3$He data.
To correct the yield when three protons were detected a similar procedure
was used, based on the angular correlation of protons from absorption
on $^3$He itself.  There was a small amount of $\pi pp$ contamination
of the selected events in this case, the magnitude of which
was determined from the spectrum of the total detected energy.

A number of corrections had to be made to the
measurement of the beam flux.  The largest
correction ($\le$~6\%) was for pions missing the target and was determined
by measuring the radial dependence of reactions in the air at both ends
of the target.  Other corrections were for beam impurity,
pion decay, and attenuation due to nuclear reactions.

To obtain the total cross section, the data had to be corrected for the
geometrical acceptance and energy threshold of the detector.
For this correction a specific model of the absorption process was
used to extrapolate over unmeasured regions.
Prior experiments \cite{Backen85,Weber91C}
have indicated that absorption on $^3$He is
apparently dominated by two distinct mechanisms:
The first is two-nucleon quasi-free
absorption (2NA), in which
the pion is absorbed on a deuteron-like pair inside the $^3$He nucleus with a
spectator proton.
The second is a three-body mechanism (3NA), characterized by a final state
distributed like three-body phase space.

Monte Carlo simulations of these
2NA and 3NA processes were used to determine for each the fraction of
the time LADS would detect two protons, and the fraction
of the time it would detect three protons.  The 2NA and 3NA
cross sections were then determined from the
experimental frequencies of these two possible results.
The weighted average of the 2NA and 3NA corrections required for the
total cross section were 15\%, 13\%, and 10\%
at 118, 162, and 239~MeV respectively.
Since these corrections are small, the determination of the total cross
section for absorption on $^3$He is not very sensitive to the
details of the models.

The determination of partial cross sections for
2NA and 3NA is more dependent on the details of the simulations.
Three-nucleon phase space was used to generate both the
2NA and the 3NA distributions.  The former was weighted so that the spectator
nucleon had a momentum distribution determined from theoretical calculations of
the $^3$He ground state wave function \cite{Ishikawa94}; these calculations
are consistent with the results of quasifree $^3$He(e,e$^{\prime}$p)
scattering \cite{Jans82,Marchand88}.  It was also weighted so that
the nucleons on which the absorption occurred had outgoing momenta and angular
distributions consistent with Ritchie's parameterization \cite{Ritchie91}
of absorption on deuterium.  The specific characteristics
of the detector were then
taken into account in order to determine its response to each simulated event.

The assumption that absorption proceeds through the two above-mentioned
mechanisms can be tested by comparing differential distributions measured by
LADS with those produced by the simulations.
Figure~2 shows the energy spectrum of the least energetic proton following
an absorption reaction.
When only two protons were detected, the energy of the third was
kinematically reconstructed, with a cut on its reconstructed mass
to select only {\it ppp} final states.
Figure~2 also shows
the results of the Monte Carlo simulations for 2NA and for 3NA, with magnitudes
normalized to the data.  The sum of the
results of the two Monte Carlos is plotted and reproduces the general
features of the data, providing support for
this model of the absorption process.
Angular distributions have also been examined, and
some deviations from phase
space in differential distributions for 3NA are indicated by the data.
However, they are not significant
for this measurement of integrated cross sections, which are determined
directly
by counting the number of absorption events with only small extrapolations
over unmeasured phase space.

The cross sections are presented in Table~I.
Also included is the total cross section for
absorption on deuterium measured by LADS, along
with the values predicted by the parameterization of reference \cite{Ritchie91}
for comparison.
There were several experimental contributions to the systematic uncertainty
of the total cross sections, with magnitudes up to 2.3\%.
These include uncertainties in the determination of the reaction losses,
MWPC efficiency, acceptance, energy threshold,
background subtraction, and normalization.
No uncertainty was estimated for the assumption that absorption proceeds
only through the 2NA and 3NA mechanisms, although uncertainties
reflecting those in the final state angular distribution and in
the initial state wave function of the absorbing pair were included.
The latter includes the question of how the absorbing $T=0$ pair differs from
a real deuteron, and it dominates the uncertainty in the 3NA cross section.

Figure~3(a) shows the LADS results for the total absorption
cross section on $^3$He
in comparison with previously reported data.  Figure~3(b) is the same for
the 2NA cross section, and Fig.~3(c) shows the 3NA cross section.
At 118~MeV the cross sections are significantly
larger than previously reported.  It is important to note the
agreement at this energy, as well as at the other two, between the
measured cross section on deuterium and
that from Ritchie's parameterization \cite{Ritchie91}.

The ratio of 2NA on $^3$He to absorption on $^2$H has been the subject of
considerable discussion \cite{Weyer90,Ingram92,Ohta85}.  The
LADS results for this ratio are 1.86 $\pm$~0.10,
1.60 $\pm$~0.09, and 1.63 $\pm$~0.15 at 118, 162, and 239~MeV
respectively.
The fact that this ratio is fairly close to 1.5 has
sometimes been interpreted as an indication that the cross section scales
simply with the number of $T=0$ ``deuteron-like'' pairs in $^3$He.
This however neglects possible effects in $^3$He such as those
due to the smaller inter-nucleon spacing, the existence of more competing
reaction channels, the larger binding energy, and initial and
final state interactions.

In conclusion, cross sections have been measured
for pion absorption on $^3$He.
The total has been divided into 2NA and 3NA components, with 22\%, 29\%, and
30\% of the total attributable to 3NA at
118, 162, and 239~MeV incident pion energy respectively.
These are the first results for the total and three-nucleon
absorption cross sections on $^3$He coming from direct measurements
without large extrapolations over unobserved regions.
The significant 3NA cross section confirms previously
reported results from detectors with much smaller solid angle coverage.
However, the 2NA component
and the total absorption cross section, while still
roughly consistent with previous results, appear to peak at a lower
energy than previously indicated.
This shift in peak energy brings observations on $^3$He into
better agreement with
those on $^2$H \cite{Ritchie91} and
on $^4$He \cite{Budagov62,Baumgartner83}.

We thank the staff of the Paul Scherrer Institute for their
considerable assistance in mounting and running this experiment.
This work was supported in part by
the German Bundesministerium f\"{u}r Forschung und Technologie,
the German Internationales B\"{u}ro der Kernforschungsanlage J\"{u}lich,
the Swiss National Science Foundation,
the US Department of Energy,
and the US National Science Foundation.

\begin{table}[p]
\begin{tabular}{lccccccc}
&&
\multicolumn{2}{c}{$^2$H} &&
\multicolumn{3}{c}{$^3$He}\\
T$_{\pi ^+}$&& LADS & Ref. \protect\cite{Ritchie91} && 2NA & 3NA & Total \\
(MeV) && (mb) & (mb) && (mb) & (mb) & (mb)\\ \hline

118 &&$11.5 \pm 0.4$& 11.9 &&$21.3 \pm 1.0$&
$6.0 \pm 0.6$&$27.3 \pm 0.8$\\

162 &&$10.9 \pm 0.3$& 10.6 &&$17.4 \pm 0.8$&
$7.2 \pm 0.7$&$24.7 \pm 0.7$\\

239 &&$4.3 \pm 0.2$& 4.2 &&$7.0 \pm 0.6$&
$3.0 \pm 0.5$&$10.0 \pm 0.4$\\
\end{tabular}
\caption
{Total cross sections for $\pi^+$-absorption
on $^2$H and
$^3$He.  Also listed are the values of Ritchie's parameterization
of absorption on $^2$H \protect\cite{Ritchie91}, and the division into
2NA and 3NA components for the absorption on $^3$He.
Both systematic and statistical uncertainties are represented;
the systematic uncertainties dominate.
\label{tableone}}
\end{table}

\begin{figure}[p]
\caption{A schematic representation of the LADS detector.
Side and axial views are shown.}
\end{figure}

\begin{figure}[p]
\caption{The energy of the least energetic outgoing proton from
the absorption of 162~MeV $\pi ^{+}$ on $^3$He.
The solid line is the data, the dotted lines are the results of the
2NA (the low energy peak) and the 3NA (the broad spectrum) simulations,
and the dashed line is the sum of the two.
The simulations include the effect of the detector's acceptance.}
\end{figure}

\begin{figure}[p]
\caption{The total absorption cross section (a) on $^3$He as well
as the 2NA (b) and 3NA (c) components, compared with previously
published data.  The solid circles are from this work, the
triangles from Aniol {\it et al}. \protect\cite{Aniol86}, the diamonds from
Mukhopadhyay {\it et al}. \protect\cite{Mukhop91},
the crosses from Smith {\it et al}.
\protect\cite{Smith89}, and the squares from
Weber {\it et al}. \protect\cite{Weber91A}.}
\end{figure}

\end{document}